# Evaluation of Account Recovery Strategies with FIDO2-based Passwordless Authentication

Johannes Kunke[1], Stephan Wiefling[1,2], Markus Ullmann[1,3], Luigi Lo Iacono[1]

**Abstract:** Threats to passwords are still very relevant due to attacks like phishing or credential stuffing. One way to solve this problem is to remove passwords completely. User studies on passwordless FIDO2 authentication using security tokens demonstrated the potential to replace passwords. However, widespread acceptance of FIDO2 depends, among other things, on how user accounts can be recovered when the security token becomes permanently unavailable. For this reason, we provide a heuristic evaluation of 12 account recovery mechanisms regarding their properties for FIDO2 passwordless authentication. Our results show that the currently used methods have many drawbacks. Some even rely on passwords, taking passwordless authentication ad absurdum. Still, our evaluation identifies promising account recovery solutions and provides recommendations for further studies.

## 1 Introduction

At least since the COVID-19 pandemic and the inevitable need for work from home solutions for employees [if20, De20], online services and the associated authentication processes have become indispensable. However, this shift in online use also increased attacks on password-based authentication. Credential stuffing, a very successful and scalable attack, where attackers automatically enter leaked login credentials (username and password) on an online service, increased in March 2020 [Ak20]. Similarly, the amount of phishing attacks increased since March 2020 [An20]. To protect against these type of attacks, multiple measures are possible. For online services with a certain amount of sensitive data, risk-based measures to strengthen password-based authentication showed to provide high security with good usability [WDLI20, WDLI21]. Another and even better approach to eliminate attacks on password-based authentication would be to replace passwords with a secure alternative. The FIDO2 Universal Authentication Framework (UAF) standard is a promising solution to implement passwordless authentication. This standard allows strong cryptographic single factor authentication and uses a different asymmetric cryptographical key for each online service [FI17]. Users can provide these keys, e.g., via a security token (also known as the authenticator). The clear advantage of FIDO2 UAF is that users do not have to construct or remember secrets. This is one of the reasons why it is considered to be more usable than classical password-based authentication in some cases [Ly20]. However, users expressed concern about losing service access if the security token becomes permanently unavailable [Ly20, Fa20]. When the token gets unavailable, access to the online services managed with it is no longer possible, since the generated secret keys never left it and thus are only available there. Complicating matters further, the security token stores different secrets for different online services. For this reason, widespread adoption of FIDO2-based passwordless authentication relies on adequate recovery mechanisms.

[1] H-BRS University of Applied Sciences, FB Informatik, Grantham-Allee 20, 53757 Sankt Augustin, Germany
johannes.kunke@smail.inf.h-brs.de, {stephan.wiefling,markus.ullmann,luigi.lo_iacono}@h-brs.de
[2] Ruhr University Bochum, Horst Görtz Institute for IT-Security, Universitätsstrasse 150, 44780 Bochum, Germany
[3] Bundesamt für Sicherheit in der Informationstechnik, Heinemannstraße 11, 53133 Bonn, Germany

**Contributions.**  We examine 12 account recovery mechanisms in passwordless FIDO2 environments using security and usability metrics evaluated in the literature. Our results suggest that widely deployed account recovery mechanisms need to be replaced with more secure alternatives. We give recommendations on promising account recovery mechanisms. Our work supports service owners and developers on which account recovery mechanism to choose for their FIDO2 implementation. It also supports researchers to select promising account recovery mechanisms for future studies. Overall, we are contributing to the shift from password-based authentication to passwordless FIDO2 authentication on the Internet.

## 2   Evaluation Criteria

To evaluate the various account recovery mechanisms, we introduce a set of criteria established in literature. These criteria are based on those developed by Bonneau et al. [Bo12], Nielsen [Ni94], Saltzer and Schröder [SS75], and Stajano [St11]. We selected the criteria based on their relatability to account recovery mechanisms and frequent citation. Some of these criteria were initially developed for user authentication purposes. However, as account recovery is also a form of authenticating users, it is comparable to classical user authentication schemes [WI20a]. This is further underlined by the fact that the process involved in account recovery mechanisms is also called *Fallback Authentication* [Ja14]. Similar to Bonneau et al. [Bo12], we divided the criteria into three categories *usability benefits*, *deployability benefits*, and *security benefits*.

### 2.1   Usability-Benefits

(**U1**) **Memorywise-Effortless** [Bo12]: The user does not have to remember an authentication secret. (**U2**) **Scalable-for-User** [Bo12]: No additional burden is introduced when using the mechanism with hundreds of services. (**U3**) **Nothing-to-Carry** [Bo12]: The user does not need to carry any additional physical item to use the recovery mechanism at any time. (**U4**) **Physically-Effortless** [Bo12]: Users do not need to perform any physical activities during the process beyond pressing a button. (**U5**) **Easy-to-Learn** [Bo12, SS75, Ni94]: The mechanism is intuitively designed and thus easy to learn. (**U6**) **Match between system and the real world** [Ni94]: The access recovery mechanism is based on real world concepts. The user can operate it intuitively because it is based on real world operations.

## 2.2 Deployability benefits

(**D1**) **Accessible** [Bo12]: Users must be able to use this mechanism, even with physical limitations. (**D2**) **Negligible-Cost-per-User** [Bo12]: The financial cost per user must be very low. (**D3**) **Browser-Compatible** [Bo12]: Mechanism can be used with any standard web browser without installing additional plugins or other software. (**D4**) **Non-Proprietary** [Bo12]: The mechanism can be used at no additional cost for royalties and are not protected by patents or other trade secrets. (**D5**) **Implemented** [St11]: The mechanism must be implemented as a practical application. It must not exist only as a theoretical concept.

## 2.3 Security Benefits

(**S1**) **Resilient-to-Physical-Observation** [Bo12]: Despite observing the user while using the mechanism, attackers fail to successfully legitimize themselves as the user. (**S2**) **Resilient-to-Targeted-Impersonation** [Bo12]: The attacker can not impersonate the user to the mechanism with background knowledge, which he may be able to obtain, e.g., via social networks. (**S3**) **Resilient-to-Internal-Observation** [Bo12]: Despite intercepting user input at participating devices, e.g., smartphone or desktop PC, it is impossible for an attacker to imitate the user. (**S4**) **Resilient-to-Leaks-from-Other-Verifiers** [Bo12]: A user uses other services that use the same or similar mechanism but whose data is made public. The attacker cannot impersonate the user with the obtained data at that service. (**S5**) **Resilient-to-Phishing** [Bo12]: Attacker is able to fake a legitimate mechanism and convince the user to use the faked version, but cannot successfully impersonate the user to the service with the resulting data. (**S6**) **Resilient-to-Theft** [Bo12]: Refers to mechanisms that require the factor possession in the form of an object as proof of legitimacy. If attackers gain possession of a user's object, they must not succeed in legitimizing themselves as the user to the mechanism. (**S7**) **No-Trusted-Third-Party** [Bo12]: The mechanism for checking the authorization of the access recovery process is not based on a third trusted party, which could have been taken over or manipulated by an attacker to become an untrusted party. (**S8**) **Requiring-Explicit-Consent** [Bo12]: The access recovery mechanism must only be performed with the user's conscious consent. It must never be started accidentally or automatically. (**S9**) **Unlinkable** [Bo12]: The information processed by this mechanism cannot be used to draw conclusions about what other services a user is using. (**S10**) **Open** [St11, SS75]: The code or at least the functionality of the mechanism must be openly accessible to everyone. (**S11**) **Work-Factor** [SS75]: The mechanism should be designed in such a way that an attacker has to invest many resources to falsely successfully legitimize against the mechanism. (**S12**) **Complete-Mediation** [SS75]: The authorization to use the mechanism must be verified every time. It is not enough to assume that the person that operates the mechanism during an open session is the legitimate user.

# 3   System Evaluation

Based on the criteria, we evaluated 12 account recovery mechanisms described in literature and partly deployed in practice (see Table 1). The mechanisms were selected based on literature research and observed mechanisms on popular online services. To cover all literature as completely as possible, we did forward snowballing based on Lyastani et al. [Ly20], and queried ACM, Springer, and IEEE publication search engines with the keywords FIDO2, account recovery, token loss, authentication, passwordless single factor, and WebAuthn. We further queried the Google search engine with the keywords FIDO2, account recovery, and device loss, to included unpublished mechanisms as well. We discuss a subset of important criteria—due to limited space—for each recovery mechanism below.

| | | Security Questions | Password | OTP | Pico | Delegated Account Recovery | FIDO2 Backup Token | Identity Card | Advanced Protection Program | Let's Authenticate | Key Copy | Online Recovery Storage | Pre-emptive Syncing |
|---|---|---|---|---|---|---|---|---|---|---|---|---|---|
| Usability | Memorywise-Effortless | ○ | ○ | ● | ● | ○ | ● | ● | ● | ○ | ● | ● | ● |
| | Scalable-for-User | ○ | ○ | ● | ● | ○ | ● | ○ | ○ | ● | ● | ● | ● |
| | Nothing-to-Carry | ● | ● | ○ | ○ | ● | ○ | ○ | ○ | ● | ○ | ○ | ○ |
| | Physically-Effortless | ● | ● | ○ | ● | ○ | ○ | ○ | ○ | ● | ● | ● | ● |
| | Easy-to-Learn | ● | ● | ● | ○ | ○ | ● | ○ | ● | ● | ● | ● | ● |
| Deployability | Match System-Real World | ● | ● | ○ | ○ | ● | ○ | ○ | ● | ○ | ● | ○ | ○ |
| | Accessible | ● | ● | ● | ○ | ● | ○ | ○ | ● | ○ | ● | ○ | ○ |
| | Negligible-Cost-per-User | ● | ● | ● | ○ | ● | ○ | ○ | ○ | ○ | ○ | ○ | ○ |
| | Browser-Compatible | ● | ● | ● | ○ | ● | ○ | ○ | ● | ○ | ○ | ○ | ○ |
| | Non-Proprietary | ● | ● | ● | ● | ● | ● | ● | ○ | ● | ● | ● | ● |
| | Implemented | ● | ● | ● | ○ | ● | ● | ○ | ● | ○ | ○ | ○ | ○ |
| Security | Resilient-Physical-Observation | ○ | ○ | ● | ● | ○ | ● | ● | ● | ○ | ● | ● | ● |
| | Resilient-Targeted-Impersonation | ○ | ● | ○ | ● | ○ | ● | ○ | ● | ● | ● | ● | ● |
| | Resilient-Internal-Observation | ○ | ○ | ● | ● | ○ | ● | ● | ● | ○ | ● | ● | ● |
| | Resilient-Leaks-from-Other-Verifiers | ○ | ○ | ● | ● | ○ | ● | ● | ● | ● | ● | ● | ● |
| | Resilient-Phishing | ○ | ○ | ○ | ● | ○ | ● | ● | ● | ○ | ● | ● | ● |
| | Resilient-Theft | ● | ● | ○ | ○ | ○ | ○ | ○ | ● | ○ | ○ | ○ | ● |
| | No-Trusted-Third-Party | ● | ● | ● | ● | ○ | ● | ○ | ● | ● | ○ | ○ | ● |
| | Requiring-Explicit-Consent | ● | ● | ● | ● | ● | ● | ● | ● | ● | ○ | ○ | ○ |
| | Unlinkable | ○ | ● | ● | ○ | ○ | ● | ● | ○ | ○ | ● | ● | ● |
| | Open | ● | ● | ● | ● | ○ | ● | ● | ● | ● | ● | ● | ● |
| | Work-Factor | ○ | ○ | ● | ● | ● | ● | ● | ○ | ● | ● | ● | ● |
| | Complete-Mediation | ● | ● | ● | ● | ● | ● | ● | ● | ● | ● | ● | ● |

● Criteria fulfilled    ○ Criteria not fulfilled    **Bold**: Deployed in account recovery practice

Tab. 1: Comparison of account recovery strategies with FIDO2-based passwordless authentication

**Security Questions.**   Security questions can be considered as a classical recovery mechanism on online services. They are entered during account registration by the user. If users lose access to their account, they can authenticate themselves to the system by answering the security questions. Typical questions are *"Mother's maiden name"*, *"Favorite sports team"*, or *"Name of first pet"*. [AJ09][Ra08]. Such security questions proved to be a weak authentication mechanism. Rabkin [Ra08] found that 12% of their studied security questions were solvable on the first attempt using social networks. However, this recovery process is *Physically-Effortless*. The user does not need to carry any physical object. There is also no additional cost for users to use this mechanism, so *Neglibile-Cost-per-User* is fulfilled. Furthermore, security questions are not resistant to *Resilient-to-Leaks-from-Other-Verifiers*. If the user entered the same question answers on another online service, that service could be compromised. With the information gained from these data breaches, the attacker can impersonate the user on the actual online service.

**Backup Password.** A backup password is stored during registration. If users lose access to their account, they can use this password to regain access. Users must actively enter the password, so it fulfills *Requiring-Explicit-Consent*. However, backup passwords are not *Scalable-for-User*, as users have to remember new passwords for each access recovery mechanism that is based on passwords. It also does not fulfill *Work-Factor*, since attackers can guess simple passwords in a short time [Pa19].

**One-Time Password (OTP).** By proving possession of the OTP, users can authenticate themselves to the system. OTPs can be generated in many ways, e.g., during registration as a list of OTPs, or time-dependent during the recovery process (time-based OTP) [Re19]. The OTP is generated when users start the mechanism, and is communicated to them via another channel. OTPs are *Memorywise-Effortless*, as users do not have to remember a secret. It is also *Browser-Compatible*, since browsers only need to provide a text field, where users enter their OTP. However, it not *Resilient-to-Phishing*, as attackers can trick users into revealing their OTP on a phishing website and replay it on the target website [Gr18, Wi20b].

**Pico.** The Pico ecosystem [St11] consists of a so-called Pico and a Pico sibling. The Pico is an authenticator device, which comes with a docking station. The docking station charges and backs up the Pico. If users lost their Pico, they can connect a new Pico to the docking station and use the Pico sibling of the lost Pico to unlock the backup and transfer the key material of that Pico to the new Pico. Since the access recovery mechanism does not depend on a trusted third party, it satisfies *No-Trusted-Third-Party*. However, users need to purchase several components to use the Pico ecosystem. Therefore, it does not fulfill *Negligible-Cost-per-User*. Also, the mechanism does not fulfill *Requiring-Explicit-Consent* due to the fact that the recovery process starts automatically when the unused Pico, connected to the docking station, and the sibling are nearby.

**Delegated Account Recovery.** Users deposit a token for account recovery at a so-called recovery provider [Fa17]. To restore access, the online service starts a request to the recovery provider. If the user successfully authenticates against the recovery provider, the provider transmits the token back to the online service and recovers access to the user account. The protocol is not *Scalable-for-User* at the moment, as users must enter a recovery provider for each online service they like to register. When the authenticator is lost, users must perform the access recovery process for each online service individually. It is also not *Easy-to-Learn*, since the meaning of authenticating to the recovery provider may not be clear to non-technical users. Beyond that, the protocol does not fulfill *Unlinkable*. The recovery provider could record the online services from which it receives requests for a particular user. Then, it could authenticate the user and restore access for the user's other online services without the user's consent.

**FIDO2 Backup Token.**    The FIDO Alliance recommends to register another FIDO2 security token in addition to the primary FIDO security token [GLS19]. When users lost their primary security token, the second security token can be used to access the account using standard FIDO2 authentication. After that, they can remove the old security token from their account and register a new primary security token. The recovery mechanism has the advantage of being *Memorywise-Effortless*. It is also *Resilient-to-Internal-Observation*, as the authentication challenge requested by the online service is solved within the security token. However, it is not *Scalable-for-User*, since users have to register the backup security token and withdraw the old security token with each online service individually.

**Identity Card.**    Citizens in Germany can use their identity card (nPA/eID) as a FIDO authenticator [ij20]. It is possible that other countries' identity cards offer this feature as well [EZ20]. Since every German citizen older than 16 years has to possess such an identity card, it would be possible as a backup authenticator variant. Following that, this recovery mechanism is *Negligible-Cost-per-User*, in contrast to classical FIDO2 backup tokens. Since the identity card is protected with a PIN by default, it is also *Resilient-to-Theft*. However, it is not *Browser-Compatible*, as users need to install special browser plugins or smartphone apps to use this mechanism. It also fails *No-Trusted-Third-Party*, since the eID server must be used in addition to the authenticator and the online service.

**Advanced Protection Program.**    Google introduced set of precautions to protect particularly vulnerable user accounts, e.g., those of political activists [Go20]. This includes securing the account with FIDO security tokens as a second factor. As a part of this program, Google also offers the access recovery option. This means that if users lose their security tokens, they can access their account using a device that is still logged in. The account recovery can not be assured *Physically-Effortless*, as the steps to be taken and the amount of required physical effort are not further specified. Therefore, it cannot be ruled out that attackers with person-specific knowledge could carry out a successful attack. Therefore, the mechanism can not reliably fulfill *Resilient-to-Target-Impersonation*. Since the open session is sufficient to restore the account, the mechanism does not fulfill *Complete-Mediation*.

**Let's Authenticate.**    The existing FIDO2 scheme is modified to address the access recovery problem [CZ19]. The mechanism provides authentication using certificates rather than public and private keys directly. To do this, a Certification Authority (CA) is introduced into the FIDO ecosystem. For registering with an online service, users need an account at the CA. If the users lose their authenticator, the CA can issue new certificates for the online services they used with the previous security token. The access recovery mechanism is not *Memorywise-Effortless*, as users must authenticate against the CA with a password after losing their authenticator. However, the mechanism is *Scalable-for-User*, because access to all online services is immediately restored using the reissued certificate. The Let's Authenticate paper [CZ19] correctly mentions that no login credentials can be phished.

However, the mechanism is still not *Resilient-to-Phishing*, as the credentials that authenticate the user to the CA could be phished.

**Key Copy.** When using a mobile device as an authenticator, the key material can be transferred from one authenticator to another [Ni18]. In doing so, the security principle that private key material must never leave the authenticator is relaxed. A so-called Owner Identitification Service (OIS) ensures that the two devices that are to exchange keys belonging to the user. The recovery mechanism is *Unlinkable*, as it still stores different secrets per online service. It is also *Browser-Compatible*, since it does not require any additional browser functionality. However, it only partially scales for users. Users must periodically perform the key copy mechanism between the two authenticators. Nevertheless, users do not need to register both authenticators on an online service every time. We still attributed *Scalable-for-User*, as users decide how often they perform the key copy mechanism. Since it relies on the OIS, however, the mechanism does not fulfill *No-Trusted-Third-Party*.

**Online Recovery Storage (ORS).** In this mechanism, the user has a primary and a backup security token, and access to a third party ORS [Ta]. The user generates a large number of keys on the backup security token, which the security token signs. The ORS stores the signed keys. During registration on a online service, the primary security token generates a new asymmetric key pair, and sends the public key to the service. Then, the primary security token decrypts a data block from the ORS. In the decrypted data, the primary authenticator adds information of its registration to the new online service. These include a unique app ID and the public key sent to the online service. The access recovery works via newly generated keys and delegations of the backup security token keys via the transfer access protocol [TKC17]. The mechanism is *Scalable-for-User*, as users only need to connect their new security token to the backup security token once to regain access to all of their accounts. Also, it is *Unlinkable*, since the mechanism stores the data block encrypted using the privacy wrapping key on the ORS. Therefore, the information is not traceable to other parties and the user is not traceable. It is not *Browser-Compatible* at the moment, as the concept integrated a non-standardized FIDO protocol message (transfer-access response).

**Pre-emptive Syncing.** The user has a primary and a backup security token in this mechanism [Ta18]. Before the primary security token is registered with an online service for the first time, both primary and the backup security token must be paired over a secure channel. The backup authenticator then generates a sufficiently large number of asymmetric key pairs. To use a new primary security token, the user initiates the Transfer Access Protocol [TKC17] between the new primary and the backup security token. The backup security token informs the new primary security token of the number of key pairs it generated at that time. Then, the new security token generates the same number of asymmetric key pairs and sends the respective public keys to the backup security token. The backup security token delegates each of its private keys to one of the public keys sent by the new security token (see Figure 1). Based on the delegations, the online service can verify that the new primary security token is authorized to access the account. The functionality of this mechanism is very similar to ORS, so the fulfilled criteria can be adopted. In addition,

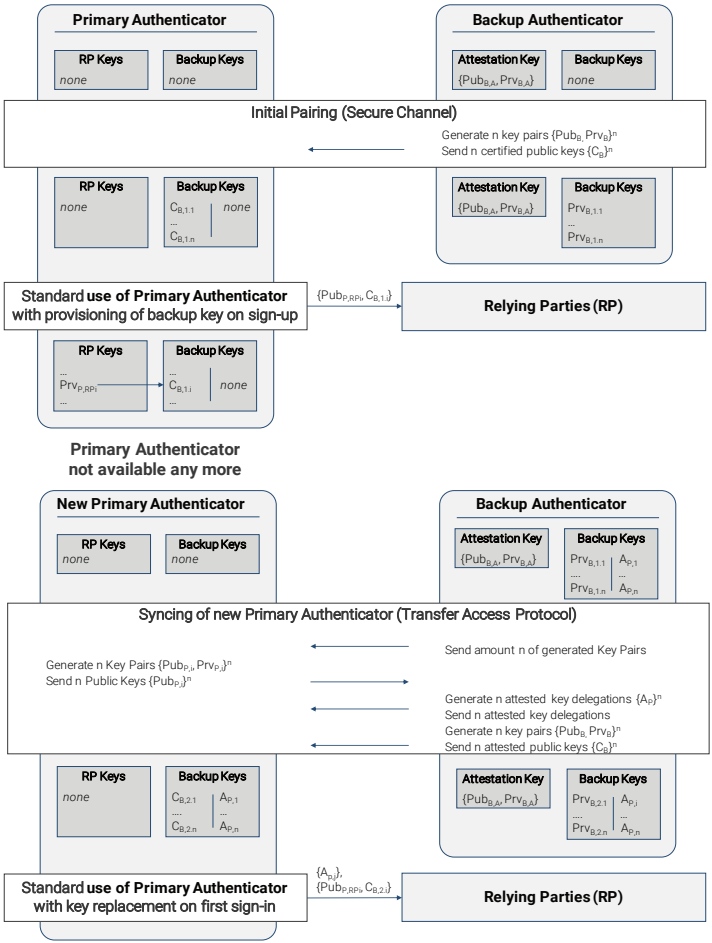

Fig. 1: Replacing a no longer available authenticator with the help of a backup authenticator without a trusted third party based on Pre-emptive Syncing

*No-Trusted-Third-Party* is fulfilled, as the mechanism only requires two security tokens and no third party. The mechanism is also *Unlinkable*, but for a different reason. Pre-emptive Syncing only exchanges the public key parts with the online services. As it does not require a third party, there is no point at which tracing users across online services would occur.

## 4   Discussion

The results confirm that security questions should be avoided at all costs. They performed significantly worse than passwords and provide an attack vector due to publicly accessible sources. We further discuss a selection of relevant recovery mechanisms based on our results and findings in the following.

**Passwords** are widely used and recognized by users. Based on the evaluated criteria, they also performed quite good. However, users need to remember passwords. Thus, using password-based recovery in a passwordless FIDO2 authentication systems would invalidate the advantages of passwordless authentication. Facebook's **Delegated Account Recovery Protocol** addresses this problem and shows an interest to solve it. However, Facebook itself also has an interest being positioned as a major identity and recovery provider in the future. This can be seen in their initiatives to provide a Single Sign-On solution for other websites [Ka18]. Access recovery using a **FIDO2 backup security token** performed among the best in comparison. Some service providers that offer passwordless FIDO2 authentication also already use this mechanism, since the FIDO alliance recommends this method. The backup authenticator variant with the Identity Card performed slightly worse than the FIDO2 backup variant. For German citizens, this would be a cost-effective alternative with regard to purchasing a second authenticator. A problem, however, is that the Identity Card service would be able to store the online services a user is registered to. Thus, we assume that users will likely reject this method on some online services. The **Online Recovery Storage** mechanism provides a key revocation mechanism. If users lost their primary security token, the ORS can automatically revoke all keys to deny access for attackers. To do this, however, the security tokens used must be able to establish a connection with the ORS. The pre-emptive syncing mechanism does not rely on connecting to a third party. From a cryptographic point of view, the positive aspect of this mechanism is that, in contrast to the key-copy mechanism, the principle that private key material never leaves the security token can still be fulfilled. The disadvantage of this mechanism is the memory and computational load that the security token have to initially carry out once to create the keys.

## 5 Limitations

The results are limited to passwordless FIDO2 single-factor authentication systems only. They can not relate to access recovery mechanisms in the context of two- or multi-factor authentication. We did a heuristic evaluation based on validated metrics in literature. We did not test the FIDO2 recovery mechanisms on real users, and testing all 12 mechanisms in a usability study would be beyond the scope of this research. Nevertheless, our results provide valuable guidance to foster further research on passwordless FIDO2 authentication.

## 6 Related Work

To the best of our knowledge, there are no studies evaluating account recovery mechanisms in terms of passwordless FIDO2 authentication. However, there is related work regarding Fallback Authentication in terms of password-based authentication. Markert et al. [Ma19]

described a study to investigate Fallback Authentication on a long-term. However, they did not show any results to date. Hang et al. [Ha15] studied personal experiences when users had to use account recovery on their smartphone. Based on their results, they recommended to provide multiple recovery mechanisms to address users who fail at one of the schemes. Some of our tested schemes could also provide this solution, e.g., either FIDO2 Backup Token or Identity Card. Ulqinaku et al. [Ul20], however, describe a social engineering attack to downgrade FIDO2 2FA. This attack is possible if users are able to choose an alternative authentication method besides the FIDO2 security token. However, this attack does not target the recovery mechanism.

## 7   Conclusion

In order to support user acceptance for passwordless FIDO2 authentication, its recovery mechanisms must be designed effectively. Therefore, we compared 12 account recovery mechanisms regarding their properties for FIDO2 passwordless authentication.

Our evaluation shows that most currently deployed recovery mechanisms performed worse in contrast to ones that only exist in theory to date (see Table 1). Also, knowledge-based access recovery mechanisms nullify many advantages achieved by FIDO2. Therefore, we do not recommend using them. In contrast to them, the FIDO2 backup token mechanism, which also the FIDO Alliance recommends, performed best. To mitigate the still significant usability weaknesses of this mechanism, the pre-emptive syncing mechanism is the most promising variant to provide FIDO2 in passwordless systems with a manageable and secure access recovery. This can achieve the best balance between security, privacy, and usability among all analyzed mechanisms.

In future work, the concept of pre-emptive syncing should be further investigated to address the problem of memory and computational load due to the pre-generated keys on the authenticators. Furthermore, the FIDO Alliance could take up the proposal to adopt the Transfer Access Protocol and the associated Transfer Access Message or concept in its standards. This would allow the two mechanisms, Online Recovery Storage and Pre-emptive Syncing, to become real-world usable mechanisms for access recovery in passwordless FIDO2 environments. In this way, the FIDO Alliance could eliminate the major problem of inadequately standardized access recovery to date and thus increase user adoption of passwordless FIDO2 authentication.